\newcommand{\be}{\begin{equation}}
\newcommand{\ee}{\end{equation}}

\newcommand{\bea}{\begin{eqnarray}}
\newcommand{\eea}{\end{eqnarray}}

\documentclass[aps,prb,twocolumn,superscriptaddress,10pt]{revtex4-1} 
\usepackage{amsmath,amssymb,graphicx,enumitem}

\begin{document}
\title{Observation of Intensity Statistics of Light Transmitted Through 3D Random Media}
\author{Tom Strudley}
\thanks{These authors contributed equally}
\affiliation{Faculty of Physical and Applied Sciences, University of Southampton, Highfield, Southampton SO17 1BJ, UK}
\author{Duygu Akbulut}
\thanks{These authors contributed equally}
\affiliation{Complex Photonic Systems (COPS),  MESA+ Institute for
Nanotechnology, University of Twente, P.O.Box 217, 7500 AE Enschede, The Netherlands}
\affiliation{Present address: ASML, Flight Forum 1900, 5657 EZ Eindhoven, The Netherlands}
\author{Willem L. Vos}
\affiliation{Complex Photonic Systems (COPS),  MESA+ Institute for
Nanotechnology, University of Twente, P.O.Box 217, 7500 AE Enschede, The Netherlands}
\author{Ad Lagendijk}
\affiliation{Complex Photonic Systems (COPS),  MESA+ Institute for
Nanotechnology, University of Twente, P.O.Box 217, 7500 AE Enschede, The Netherlands}
\author{Allard P. Mosk}
\affiliation{Complex Photonic Systems (COPS),  MESA+ Institute for
Nanotechnology, University of Twente, P.O.Box 217, 7500 AE Enschede, The Netherlands}
\author{Otto L. Muskens}
\email{Corresponding author: O.Muskens@soton.ac.uk}
\affiliation{Faculty of Physical and Applied Sciences, University of Southampton, Highfield, Southampton SO17 1BJ, UK}
\begin{abstract}
We experimentally observe the spatial intensity statistics of light transmitted through
 three-dimensional isotropic scattering media. The intensity distributions measured through layers 
 consisting of zinc oxide nanoparticles differ significantly from the usual Rayleigh statistics associated with speckle, and instead
 are in agreement with the predictions of mesoscopic transport theory, taking into account the known material parameters of the samples.
  Consistent with the measured spatial intensity fluctuations, the total transmission fluctuates. 
The magnitude of the fluctuations in the total transmission is smaller than expected on the basis of quasi-one-dimensional (1D) transport theory, which indicates that  
quasi-1D theories cannot fully describe these open three-dimensional media.
\end{abstract}
%
\maketitle 
Understanding the flow of light in three-dimensional (3D) scattering
environments is important for a variety of
applications ranging from new developments in biomedical imaging \cite{Park13} to
energy harvesting \cite{Vynck12}, spectroscopy \cite{Cao13,Mazilu14}, information control
\cite{Mosk12} and lighting \cite{Leung14}. Often, light transport in such media can be approximated as a series of uncorrelated, random
events. However,  interference between scattered fields can give rise to mesoscopic effects that can
reveal valuable information on the properties of the medium and the transport process.
In quantum transport theory for electrons, mesoscopic correlations originate from
the crossing of many possible trajectories inside the medium \cite{akkermansbook}. The analogy between electrons and matter waves with classical waves such as light and sound
allows a direct mapping of concepts from mesoscopic transport theory. 
The crossing probabilities of wave paths and the resulting
correlations are described by a single parameter, the
dimensionless conductance $g$, which is equal to the average number of
open transmission eigenchannels. In a waveguide geometry, $g$ is
defined as $g = N \langle T_{a}\rangle$, with $N$ the total number of
transmission eigenchannels that light in the incident free-space
modes can couple to and  $\langle T_a\rangle \approx l_{\mathrm{tr}}/L $ is the ensemble averaged transmission probability of light, with  $l_{\mathrm{tr}}$  the transport mean free
path and $L$ the thickness.

Statistical methods have been widely employed in the study of wave
transport through disordered systems to extract
mesoscopic transport contributions~\cite{Garcia1989, deBoer1994PRL, Stoytchev1997aa,
Scheffold1997, Scheffold1998, Chabanov2000,Pena14,
Zhang2002,
Page2008, Smolka2011,
 Strudley2013}. The dimensionality of the experiment is
of paramount importance. Quasi one-dimensional (1D) waveguides allow the direct
observation of light propagation in the strongly mesoscopic regime,
including Anderson localization \cite{Chabanov2000,
Pena14}. In a study conducted with visible light
using stacks of glass slides, deviations from Rayleigh statistics
were observed in the crossover from the 1D to the quasi-1D
regime~\cite{Park2010}. In a 2D study using
near-infrared light a deviation from Rayleigh
statistics was observed \cite{Riboli2011}. Of interest also are observations of freak wave phenomena in quasi 2D resonators~\cite{Hohmann2010}. The case of 3D random media is of special interest, as only in 3D a phase transition to localization is expected. In 3D materials however, the effects of 
interference are generally much weaker because the large avaliable phase
space leads to a reduced probability for trajectories to
cross. Mesoscopic effects for light in 3D media are therefore generally quite subtle and difficult
to measure
\cite{deBoer1992,deBoer1994PRL,Scheffold1997,Scheffold1998}.
Intensity statistics deviating from Rayleigh statistics in 3D have only been observed
in strongly anisotropic disordered mats of semiconductor nanowires
\cite{Strudley2013}. In the intensity statistics of light transmitted through isotropic 3D media no deviations from Rayleigh statistics have been observed  to date.
The observation of transmission statistics beyond the Rayleigh regime is a crucial test
for the extension of mesoscopic transport theory to these inherently open 3D media. 

\begin{figure}[t]
\centering
\includegraphics[width=.8\columnwidth]{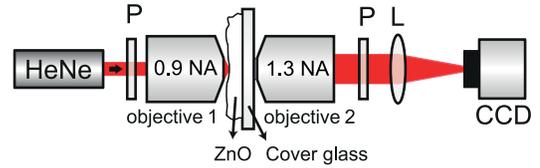}
\caption{ Experimental setup. HeNe: laser (=$\lambda=$632.8~nm, P= 5 mW). ZnO: sample. Objective
1: 100$\times$ 0.9-N$\!$A objective. Objective 2: 100$\times$
1.3-N$\!$A oil immersion objective. L: 200~mm focal length lens.
P: polarizer. CCD: camera sensor. } \label{fig:setup_specklestat}
\end{figure}

Here we present measurements of the intensity fluctuations in the transmission of light through 3D layers of ZnO particles with an average size of 200 nm. Transmission was recorded spatially using a high numerical aperture (NA) transmission microscope as shown in Fig.~\ref{fig:setup_specklestat} and described in more detail in Ref.~\cite{Strudley2013}. Light from a Helium-Neon laser  was focused on the incident surface of the sample. The spatial distribution of transmitted light at ZnO-glass interface was imaged onto a camera, in a cross-polarized configuration. 

\begin{figure*}[t]
\centering
\includegraphics[width=1.6\columnwidth]{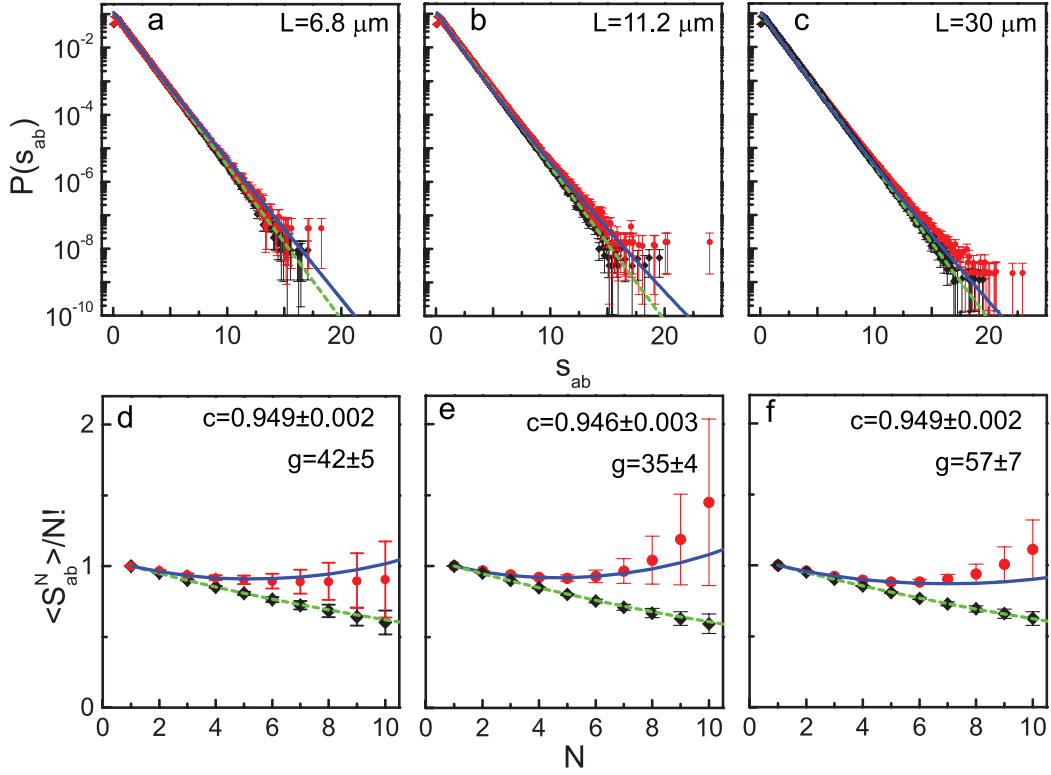}
\caption{ [wide figure] (a-c) Histograms of the intensity distribution $P(s_{ab})$ of
fields transmitted through samples of thickness $L=$ (a) 6~$\mu$m, (b) 11~$\mu$m , and (c) 30~$\mu$m.
Red and black data points: Mean of normalized histograms of six
different datasets captured IF and  OF,
respectively. Error bars: standard error of the normalized
histograms. Dashed line, green: Rayleigh statistics with reduced
contrast of $c=$0.95. Solid line, blue: Plot of Eq.~\ref{eq:specklestatNieuwenhuizen}  with $g$ from fits of (d-f). (d-f) Moments of the intensity distributions of transmitted fields for IF (red dots) and OF (black diamonds) configurations. Dashed line, green: Fits of first 5 OF moments using Rayleigh theory with $c=0.949 \pm 0.002$ (d,f) and $c=0.946$ (c). Solid line, blue: Fits of IF moments with Eq.~\ref{eq:moments} for $g=42 \pm 5$ (d), $g=35 \pm 4$ (e) and $g=57 \pm 7$ (f).} \label{fig:11pt25umand30pt7umhistograms}
\end{figure*}
We performed measurements in two configurations, `in-focus' (IF) and `out-of-focus' (OF) illumination, corresponding to illumination spots with a width at $1/e$  of less than 0.5~$\mu$m  and 25~$\mu$m, respectively.
In each configuration, 1000 images were captured per dataset, translating the sample over 1~$\mu$m for each consecutive image. At this distance, the speckle patterns of any two consecutive images were found to be completely different. A total number of 6 datasets per sample
were recorded in a procedure identical to that followed in Ref.~\cite{Strudley2013}, in brief:
 For each dataset, the captured images were averaged to obtain an
average spatial intensity distribution. The total
transmitted intensity for each sample position was obtained by
summing the total intensity in the corresponding image. In order
to divide out any sample variations over long length scales, the
total transmission was normalized to a moving average over
10~$\mu$m. This distance is shorter than the typical sample variations, and
longer than the mesoscopic correlation length in the system. A
constant background 
corresponding to the dark counts of the camera was subtracted from all captured images. Finally, each
image was cropped to the area of interest and divided pixel by
pixel by the average intensity image to obtain the normalized
intensity $s_{ab}=T_{ab}/\langle T_{ab}\rangle$, where $a$ denotes an input mode and $b$  an output mode, and the brackets denote ensemble averaging. This procedure
divides out the envelope in the intensity due to diffusion. All intensities for the 6 datasets were collected into a single
histogram, to obtain $P(s_{ab})$ vs $s_{ab}$. 
After collection of the IF histogram, the illumination objective was
moved out of focus by 25~$\mu$m and the measurement was
repeated in this `out-of-focus' (OF) configuration. The number
of transmission channels addressed by the incident field is
large when OF and small when IF. Thus, the measurement made when
OF serves as a reference for the case of large $g$ and negligible
mesoscopic corrections, whereas the measurements performed when IF
are expected to give rise to 
strong mesoscopic fluctuations.

In Fig.~\ref{fig:11pt25umand30pt7umhistograms} we show the spatial
intensity histograms $P(s_{ab})$ vs $s_{ab}$ using the data
captured IF and OF for the samples of $L=$ 6 $\mu$m (a), 11 $\mu$m (b) and 30 $\mu$m (c). 
The OF data is expected to follow Rayleigh statistics, $P(s_{ab})=\exp(-{s_{ab}})$, however due to experimental aberrations the tail of the OF distribution is slightly 
suppressed and 
 it is described by $P(s_{ab})=\exp{(-{s_{ab}}/{c})}$,
where the speckle contrast parameter $c=0.95$ models the experimental reduction in contrast.
The histograms of the data captured IF show a heavy tail at high speckle intensities, which is not present in the data captured OF 
and therefore is a direct signal of mesoscopic fluctuations.

\begin{figure}[h]
\centering
\includegraphics[width=0.65\columnwidth]{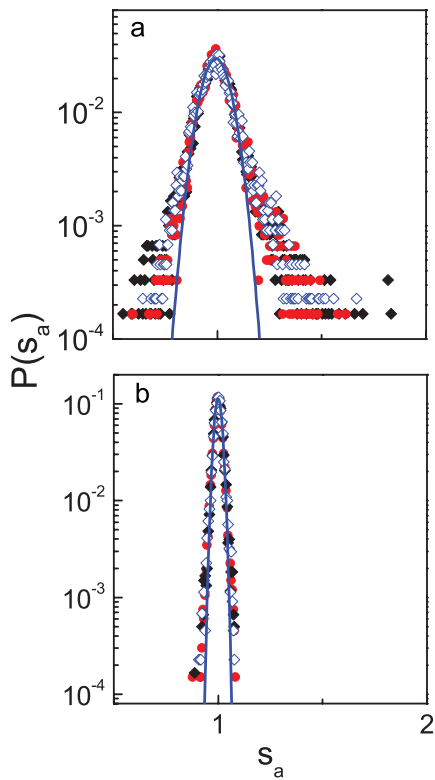}
\caption{Histograms of the total transmission $s_a$ for (a) IF and (b) OF, for ZnO samples of thickness $L=$6~$\mu$m (open diamonds, blue), 11~$\mu$m (dots, red) and 30~$\mu$m (diamonds, black). Lines: Gaussian fits. Variances of distributions are shown in Table~\ref{table:I}.} \label{fig:sa}
\end{figure}

We compared the histogram of $s_{ab}$ in
Fig.~\ref{fig:11pt25umand30pt7umhistograms} for IF with a
theoretical model for the mesoscopic distribution, given by
~\cite{Schnerb91,Nieuwenhuizen1995,Kogan1993aa,remark1}
\be
P(s_{ab})=\exp{(-\frac{s_{ab}}{c})}\left\{1+\frac{1}{3g}\left(\left(\frac{s_{ab}}{c}\right)^{2}-4\frac{s_{ab}}{c}+2\right)\right\}.
\label{eq:specklestatNieuwenhuizen} 
\ee
A robust method to analyze distributions is to fit their moments. An
analytical expression of the moments of the intensity distribution was developed
by Kogan \emph{et al.}~\cite{Kogan1993aa, remark1} and reads
\be \label{eq:moments}
<s_{ab}^N>=N!\left(\frac{<s_{ab}>c}{2ig^{1/2}}\right)^N H_N(ig^{1/2}),
 \ee
with $H_N$ the Hermite polynomial of order $N$. 
The moments of the measured distributions are shown in Fig.~\ref{fig:11pt25umand30pt7umhistograms}d-f. We see a large difference between the OF (black diamonds) and IF (red dots) moments.
The analytical
expression (\ref{eq:moments}) was fitted to the first five
moments of the data, using the contrast $c$ and the dimensionless
conductance $g$ as the free parameters.
The fits to the first 5 moments of
the OF histogram resulted in $g$ values on the order of $10^{5}$,
indicating that we can not distinguish them from a Rayleigh distribution (infinite $g$). We use the OF fits to obtain the contrast values $c$ with values indicated in the figure. For moments $N>6$, the statistical variations due to rare events give rise to a large uncertainty and we did not include these moments in the fits. 
The first 5 moments of the IF distribution were fit using only $g$ as a free parameter, with the experimental contrast parameter $c$ fixed by the OF data.
Best fits
 (shown in Fig.~\ref{fig:11pt25umand30pt7umhistograms}d-f) rwere found for  $g=42 \pm 5$ for the 6-$\mu$m sample, $g=35 \pm 4$ for the 11-$\mu$m sample, and $g=57 \pm 7$ for the 30-$\mu$m sample. The fits agree with the data up to the 10$^{\rm th}$ moment, indicating that mescosopic theory describes the observed intensity fluctuations well.

\begin{table}[t]
\begin{tabular}
{c c c c} ZnO thickness & $\frac{{\rm var}(s_{ab}^{\rm IF})}{{\rm
var}(s_{ab}^{\rm OF})}-1$   & ${\rm var}(s_{a}^{\rm IF})$ &
${\rm var}(s_{a}^{\rm OF})$  \\
\hline
6 $\mu$m & $3.2 \times 10^{-2}$ & $9.1 \times 10^{-3}$ & $3.6 \times 10^{-4}$\\
11 $\mu$m & $3.7 \times 10^{-2}$ & $8.1 \times 10^{-3}$ & $3.9 \times 10^{-4}$\\
30 $\mu$m & $2.0 \times 10^{-2}$ & $1.0 \times 10^{-2}$ & $3.7 \times 10^{-4}$\\
\hline
\end{tabular}
\caption{\label{table:I} Measured  variances of the intensity distributions.}
\end{table}

A 3D medium differs from a waveguide in the fact
that energy can spread out
in the transverse direction. 
This geometry has been modeled as a waveguide of expanding width \cite{Scheffold1997,Scheffold1998} with an effective conductance parameter $g$ that increases with $N$ and $l_{\mathrm{tr}}$, but
saturates with increasing sample thickness $L$.
In the case of an incident beam that is tightly focussed on the sample
the expanding waveguide model predicts a conductance
 $g=(8l_{tr}k^{2}/15)w_{\mathrm{min}}$.
Here $k$ represents the wavevector inside the medium 
and $w_{\mathrm{min}}=a\,l_{\mathrm{tr}}$ is the minimum width of the incident spot
inside the sample, where $a$ is a constant expected to be close to unity. A later elegant approach \cite{GarciaMartin2002} obtains similar results.
We use the previously measured values
$l_{\mathrm{tr}}$=0.7$\pm$0.2~$\mu$m and
$n_{\mathrm{eff}}$=1.4$\pm$0.1~\cite{Vellekoop2008aa} 
in the  expanding waveguide formalism and
find $g$ in the range $(50.47 \pm 29.73)a$, in good agreement with the fits to the histograms.

In addition to the spatial intensity statistics, our experimental configuration provides measurements of the total transmitted intensity. For this purpose, we integrated the total counts in each camera image, which we normalized to the ensemble average to obtain the normalized total transmission $s_a$. Figure~\ref{fig:sa} shows the IF  and OF distributions $P(s_a)$ for the two samples under study. The total transmission is the sum of a large number of independent speckle spots, and the exponential distribution converts to a Gaussian with a variance inversely proportional to the number of independent transmission channels $g$. The OF data shows a correspondingly narrow distribution, which is very similar for the three samples under study. The IF data showed a much broader distribution indicating a reduction of the number of contributing `open' channels $g$, again with little variation between the three samples.
The measured variances for both the $s_a$ and the $s_{ab}$ distributions are summarized in Table~\ref{table:I}. For the latter we corrected the IF variance for the finite speckle contrast by normalizing to the OF variance.
For the quasi-1D geometry of a waveguide, random matrix theory predicts 
 the relation ${\rm var} s_{ab}=1+2 {\rm var}(s_a)$ \cite{Chabanov2000}. It is unknown whether this theoretical
 relation can be extended in unmodified form for vector waves in 3D media. Table~\ref{table:I} shows that the measured variance of the total transmission ${\rm var}(s_{a})$ is smaller than expected from quasi-1D random matrix theory for all but the thickest sample. The fact that this deviation appears for an isotropic 3D medium as well as for highly anisotropic nanowires ~\cite{Strudley2013} indicates that the quasi-1D theory cannot generally be applied to 3D geometries.

In conclusion, we have presented measurements of the intensity statistics of light transmitted through three-dimensional isotropic ZnO
scattering media. The results show deviations from Rayleigh statistics. 
Using an analysis
of the moments of the distribution we obtain values of the dimensionless conductance $g$ of around 40, in agreement with predictions based on previously reported
sample parameters. 
This is the first direct observation through intensity statistics of
strong mesoscopic effects of light transmitted through isotropically scattering 3D  samples. 
Our results show a good agreement of the modes of the intensity distribution with transport theory. However, the ratio between the variances of the total transmission and the speckle transmission differs from the predictions of quasi-1D theory.  
Ultimately, mesoscopic effects affect important applications such as wavefront shaping and time reversal methods \cite{Mosk12},
and spectroscopy \cite{Cao13, Mazilu14}. Our results show that the regime where mesoscopic fluctuations are measurable is accessible using conventional scattering materials, opening up new avenues for experimental investigation.

This work is supported by FOM-NWO, FOM program ``Stirring of light'' and ERC grant 279248. O.M. acknowledges support by EPSRC through fellowship EP/J016918/1.


\begin{thebibliography}{99}
\bibitem{remark1}  Equations~\ref{eq:specklestatNieuwenhuizen} and \ref{eq:moments} were adapted by us to account for the experimental speckle contrast parameter $c$.
\end{thebibliography}
\end{document}